\documentclass[12pt]{article}
\usepackage{amsmath,amsfonts,amssymb}
\usepackage{graphics}
\makeatletter \@addtoreset{equation}{section}

\makeatletter\renewcommand\section{\@startsection {section}{1}{\z@}%
                                   {-3.5ex \@plus -1ex \@minus -.2ex}
                                   {2.3ex \@plus.2ex}%
                                   {\normalfont\large\bfseries}}
\renewcommand\subsection{\@startsection{subsection}{2}{\z@}%
                                     {-3.25ex\@plus -1ex \@minus -.2ex}%
                                     {1.5ex \@plus .2ex}%
                                     {\normalfont\bfseries}}

\def\baselinestretch{1.2}
\newcommand{\be}{\begin{equation}}
\newcommand{\ee}{\end{equation}}
\newcommand{\beq}{\begin{eqnarray}}
\newcommand{\eeq}{\end{eqnarray}}

\newcommand{\bbibitem}[1]{\bibitem{#1}\marginpar{#1}}

\def\Label#1{\label{#1}%
  \smash{\hbox to0pt{\raise1ex\hbox{\tiny[#1]}\hss}}}
\def\noLabels{\let\Label=\label}
\def\nobbibitem{\let\bbibitem=\bibitem}

\def\CW{{\cal W}}

\newcommand{\RR}{\mathbb{R}}

\newcommand{\ZZ}{\mathbb{Z}}
\newcommand{\Z}{\ZZ}

\begin{document}

\begin{titlepage}

\vfil\

\begin{center}

{\Large{\bf Higher Spin Theories in AdS$_3$ and a Gravitational Exclusion Principle }}

\vspace{3mm}

 Alejandra Castro\footnote{e-mail: acastro@physics.mcgill.ca}$^{a}$, 
 Arnaud Lepage-Jutier\footnote{e-mail: arnaud.lepage-jutier@mail.mcgill.ca}$^{a}$ \&
Alexander Maloney\footnote{e-mail: maloney@physics.mcgill.ca}$^{a}$
\\

\vspace{8mm}

\bigskip\medskip
\smallskip\centerline{$^a$ \it 
McGill Physics Department, 3600 rue University, Montreal, QC H3A 2T8, Canada}
\medskip
\vfil

\end{center}
\setcounter{footnote}{0}
\begin{abstract}
\noindent  

We consider theories of three dimensional quantum gravity in Anti-de Sitter space which possess massless higher-spin gauge symmetry.  The perturbative spectrum of the theory includes higher spin excitations which can be organized into vacuum representations of the $\CW_N$ algebra; these are higher spin versions of the boundary gravitons.  We describe a fundamental bound which relates the value of the cosmological constant to the amount of gauge symmetry present.  In the dual CFT language, this is the statement that modular invariance implies that the theory can not be quantized unless the central charge is sufficiently large, i.e. if $c\ge N-1$.  This bound relies on the assumption that all of the perturbative excitations exist as full states in the quantum theory, and can be circumvented if the theory possesses a linearization instability.  The $\CW_N$ minimal models -- recently conjectured to be dual to certain higher spin AdS theories by Gaberdiel and Gopakumar -- provide an example of this phenomenon.
This result can be regarded as an example of a ``gravitational exclusion principle" in Anti-de Sitter space, where a non-perturbative quantum gravity mechanism involving black holes places a limit on the number of light degrees of freedom present.

\end{abstract}
\vspace{0.5in}

\end{titlepage}
\renewcommand{\baselinestretch}{1.05}  

\newpage
\tableofcontents

\section{Introduction}

Three dimensional quantum gravity has proven a useful testing ground for many of our ideas and conjectures concerning the microscopic nature of gravity.
One of the most interesting and important conjectures is the proposal that quantum gravity places a fundamental limit on the number of light degrees of freedom present.  This conjecture is most commonly discussed in the context of black hole entropy, where it was observed that a large number of light species of identical particles would violate holographic entropy bounds \cite{Bekenstein:1993dz,'tHooft:1993gx}.  However, this notion has surfaced in a variety of different guises over the last several decades (for example in \cite{Susskind:1995da, Bousso:2002ju,ArkaniHamed:2006dz}).  The goal of this paper is to describe a specific three dimensional scenario where this idea can be put to the test using the precision techniques of AdS/CFT.


We will focus on the case of three dimensional gravity in asymptotically Anti-de Sitter (AdS) space, and consider theories with massless higher spin gauge fields.  These theories possess a large symmetry group which can be regarded as an enhanced version of the conformal symmetry present in every asymptotically AdS theory of gravity.  The states  organize into representations of this enhanced symmetry group, hence these theories contain a large number of light degrees of freedom.  In the context of AdS/CFT, the inclusion of higher spin fields is interesting in its own right.
In string theory realizations of AdS/CFT, an infinite tower of massless higher spin fields is expected to emerge when the AdS radius becomes small \cite{HaggiMani:2000ru,Sundborg:2000wp,Konstein:2000bi,Polyakov:2001af,Mikhailov:2002bp,Segal:2002gd}.  The study of such higher spin fields should therefore be regarded as a first step in the study of quantum gravity in AdS beyond the supergravity regime.  

We note that the construction of theories of massless higher spin fields is a notoriously delicate procedure.  In four space-time dimensions, a consistent theory with an infinite tower of interacting higher spin fields was constructed by Vasiliev \cite{Fradkin:1986qy,Fradkin:1987ks,Vasiliev:1990tk,Vasiliev:1992av} (see also the recent progress of \cite{Sezgin:2002rt, Klebanov:2002ja,Giombi:2009wh,Giombi:2010vg}). 
In three space-time dimensions the story is somewhat more straightforward.  A simple class of massless higher spin theories in AdS${}_3$ can be formulated using Chern-Simons theory \cite{Blencowe:1988gj,Bergshoeff:1989ns}.  Unlike Vasiliev's four dimensional theory, which can be formulated only when there are an infinite number of higher spin fields, this theory exists for both a finite and an infinite number of higher spin fields \cite{Aragone:1983sz,Campoleoni:2010zq}. These theories describe massless higher spin gauge fields which possess no local degrees of freedom and can be regarded as higher spin versions of the graviton, which itself has no local degrees of freedom in three dimensions.  

Despite the fact that these theories have no local degrees of freedom, they have interesting quantum properties which can be understood rather precisely.  There are ``non-local" degrees of freedom which are associated with boundary excitations of the fields, generalizing the classical results of Brown and Henneaux \cite{Brown:1986nw}.  In particular, the algebra of the asymptotic symmetry group  is enlarged from two copies of the Virasoro algebra to two copies of the ${\cal W}_N$ algebra, where $N$ is the highest allowed spin  \cite{Henneaux:2010xg,Campoleoni:2010zq}. The central charge of the dual CFT can be computed, and remarkably remains unaffected by the presence of the higher spin fields. A non-trivial check of this story was provided by \cite{Gaberdiel:2010ar}, who computed the one-loop determinant of the gravitational theory and showed that it is precisely the vacuum character of  ${\cal W}_N$.

Here we will investigate the effect of these higher spin fields on the spectrum of the theory. Classically and at the linearized level the theories seem to be well defined and free of pathologies.  We would like to ask what happens once quantum effects are taken into account. Our primary tool will be the AdS/CFT correspondence, which states that to every theory of gravity in asymptotically AdS space there is a dual CFT.  Thus the structure of the theory is constrained by conformal invariance. In particular, modular invariance -- invariance under large conformal transformations in Euclidean signature -- allows us to determine the spectrum of the theory at high energies.  This gives Cardy's formula, which determines the rate of growth for the density of states at high energies.   

The basic observation of this note is a simple one.  When the value of $N$ is sufficiently large, the number of linearized states in the bulk theory -- the number of higher spin versions of boundary gravitons -- exceeds this upper bound set by Cardy's formula.  In order to prevent this we must require that 
\be\label{eqq}
{N-1} \le c = {3 \ell \over 2G_3}~.
\ee
Here we have used the Brown-Henneaux expression for the central charge of the theory in terms of the AdS radius $\ell$ and Newton constant $G_3$.
Thus the existence of a dual CFT, along with the existence of these boundary excitations, provides a bound on the amount of higher spin gauge symmetry present.  An important feature of this result is that when $N$ is large it applies to theories in the semiclassical ($\ell\gg G_3$) regime.   This can be regarded as a ``gravitational exclusion principle," where quantum gravitational effects place an upper bound on the number of light states in the theory.

We note that this bound appears only when non-perturbative effects are included, and that the classical theories discussed above appear to be free of pathology for every value of $c$ and $N$.  
It is interesting then to ask exactly what happens when we try to quantize a theory with values of $N$ and $c$ which violate the bound (\ref{eqq}).
One of two things must occur.  The first possibility is that the value of $\ell$ (or $G_3$) will be renormalized by quantum effects so that (\ref{eqq}) is satisfied. In effect, quantum corrections will drive the value of the cosmological constant towards zero to accommodate the large number of degrees of freedom.  The second possibility is that some of the dangerous perturbative states are removed from the spectrum upon quantization. This would mean that the theory has a linearization instability; apparently innocuous perturbative states are not in fact linearizations of true states in the Hilbert space.  Roughly speaking, these perturbative states are removed to accommodate the finite size of Anti-de Sitter space.  It appears that both of these possibilities can be realized in theories of AdS quantum gravity.  To see this, we will consider a simple set of CFTs with $\CW_N$ symmetry, namely the $\CW_N$ minimal models, whose bulk duals were recently discussed in \cite{Gaberdiel:2010pz}.

Finally, we wish to emphasize the intimate connection between the bound \eqref{eqq} and the physics of asymptotically AdS black holes.  Every classical theory of AdS${}_3$ gravity possesses black hole solutions, the BTZ black holes.  The Bekenstein-Hawking entropy of these black holes is precisely given by Cardy's formula for the asymptotic density of states.  Thus the bound \eqref{eqq} reflects the fact that black holes dominate the spectrum of the theory at high energy.  Indeed, we will see that there is a precise sense in which those CFTs which violate the bound \eqref{eqq} -- such as the $\CW_N$ minimal models -- are dual to theories where the spectrum of high energy states is not dominated by black holes with large area.

In the next section we will review a few salient features of $\CW_N$ symmetry and higher spin theories in AdS${}_3$.  In section 3 we discuss the bound \eqref{eqq} and its application in both the finite $N$ and $N\to\infty$ case.  In section 4 we comment on the specific realizations of these conjectures in the $\CW_N$ minimal models.  In an appendix we describe the asymptotic properties of $\CW_N$ and $\CW_\infty$ vacuum characters.

\section{Higher spin fields in AdS$_3$}

In this section we summarize the main results of \cite{Henneaux:2010xg,Campoleoni:2010zq, Gaberdiel:2010ar} concerning higher spin theories in AdS$_3$. 

Classical three dimensional general relativity with a negative cosmological constant can be rewritten as a Chern-Simons gauge theory with gauge group $SO(2,2)\simeq SL(2,\RR)\times SL(2,\RR)$ \cite{Achucarro:1987vz,Witten:1988hc,Achucarro:1989gm}.  It is easy to generalize this to include a theory with fields of up to spin $N$.  We simply replace the $SL(2,\RR)$ gauge group by $SL(N,\RR)$ \cite{Campoleoni:2010zq}. In this case the higher spin fields are massless and have no local propagating degrees of freedom; the theory is topological, just as with the spin $2$ graviton case.  Further, one can take the infinite dimensional extension of $SL(2,\RR)$ -- denoted $hs(1,1)$ -- which will describe a infinite tower of spins in a similar spirit as the Fradkin-Vasiliev theory \cite{Blencowe:1988gj,Bergshoeff:1989ns}.

To formulate this theory more precisely, we introduce a pair of tensor-valued one forms
\be\label{sec2:ax}
e_\mu^{~a_1\cdots a_{s-1}}~,\quad \omega_\mu^{~a_1\cdots a_{s-1}}~,
\ee
where $a_i$ are Lorentz indices. If the gauge group is $SL(2,\RR)$ we identify $e_\mu^{~a}$ with the dreibein and $\omega_{\mu}^{~a}$ with the spin connection. The Chern-Simons gauge fields are the linear combinations
\be
A^{\pm}_{(2)}=J_a\left(\omega_{\mu}^{~a}\pm {1\over \ell}e_\mu^{~a}\right)dx^\mu~,
\ee
where $J_a$ are the generators of $sl(2,\RR)$. The equations of motion are found by extremizing the Chern-Simons action
\be
I_{CS}[A]={k\over 4\pi}\int {\rm tr}(A\wedge dA+{2\over 3}A\wedge A\wedge A)~,
\ee
where ${\rm tr}$ is the symmetric bilinear form on $SL(2,\RR)$. The Einstein-Hilbert action is
given by
\be\label{sec2:ba}
I_{EH}=I_{CS}[A^+_{(2)}]-I_{CS}[A^-_{(2)}]~,\quad k={\ell\over 4G_3}~.
\ee
where $\ell$ is the AdS$_3$ radius and $G_3$ Newton's constant. 

To include the dynamics of the spin $s$ field, we define
\beq
A^{+}&=& A_{(2)}^++T_{a_1\cdots a_{s-1}}\left(\omega_\mu^{~a_1\cdots a_{s-1}}+{1\over \ell}e_\mu^{~a_1\cdots a_{s-1}}\right)dx^\mu~,\cr
A^{-}&=& A_{(2)}^-+T_{a_1\cdots a_{s-1}}\left(\omega_\mu^{~a_1\cdots a_{s-1}}- {1\over \ell}e_\mu^{~a_1\cdots a_{s-1}}\right)dx^\mu~,
\eeq
with $s> 2$ and $T_{a_1\cdots a_{s-1}}$ are generators of the extended gauge group.  
We can then identify the gauge fields \eqref{sec2:ax} with higher spin fields as defined by Fronsdal \cite{Fronsdal:1978rb} provided the generators $T_{a_1\cdots a_{s-1}}$ obey the correct algebra. First, the generators $T_{a_1\cdots a_{s-1}}$ must be taken to be symmetric and traceless.  Second, the $J_a$ and $T_{a_1\cdots a_{s-1}}$ must form the Lie algebra 
\be
[J_a,J_b]=\epsilon_{abc}J^c~,\quad [J_a,T_{a_1\cdots a_{s-1}}]=\epsilon^m_{~~a(a_1}T_{a_2\cdots a_{s-1})m}~.
\ee
One can then consider the Chern-Simons action
\be\label{sec2:ba}
I_{N}=I_{CS}[A^+]-I_{CS}[A^-]~.
\ee
For $N>2$, one can check that the linearized fluctuations of the gauge fields around a fixed metric background should satisfy the equations of motion of higher spin fields. More precisely, we have 
\be
e^{~a}_\mu=e^{(0)a}_\mu+e^{(1)a}_\mu~,\quad \omega^{~a}_\mu=\omega^{(0)a}_\mu+\omega^{(1)a}_\mu~,
\ee
where the upper script $(0)$ denotes the background and $(1)$ are fluctuations. Treating all other higher spin fields as fluctuations, the linearized Chern-Simons equations are reduced to 
 \be\label{sec2:bb}
 \nabla^2\varphi_{\mu_1\cdots\mu_s}^{}-\nabla_{(\mu_1|}^{}\nabla^\lambda\varphi^{}_{|\mu_2\cdots\mu_s)\lambda}+\nabla_{(\mu_1}^{}\nabla_{\mu_2}^{}\varphi_{\mu_3\cdots\mu_s)\lambda}^{\phantom{\mu_2\cdots\mu_s)\lambda}\lambda}=0~,
 \ee
where 
\be\label{sec2:bd}
\varphi_{\mu_1\cdots\mu_s}={1\over s}e^{(0)a_1}_{(\mu_1}\cdots e^{(0)a_{s-1}}_{\mu_{s-1}}e_{\mu_s)a_1\cdots a_{s-1}}~,
\ee
for $s\ge2$. At this level the connection $ \omega_\mu^{~a_1\cdots a_{s-1}}$ becomes an auxiliary field. Equation \eqref{sec2:bb} is exactly the equation of motion for a free spin $s$ field propagating on a curved space-time.

There are two very interesting results for these higher spin theories in AdS$_3$. First, in  \cite{Henneaux:2010xg,Campoleoni:2010zq} the authors computed the asymptotic symmetries of the $SL(N,\RR)\times SL(N,\RR)$ Chern-Simons theories for a given set of boundary conditions. Taking the connection $A^{\pm}_{(2)}$ on empty AdS$_3$ as the definition of ``asymptotically AdS configurations'', they found all gauge transformations that left the connection invariant up to a constant term with respect to AdS$_3$ near the boundary. The remarkable result is that the algebra of the asymptotic symmetries is given by two copies of the ${\CW}_N$ algebra. Further, the algebra allows for a central extension and its central charge is 
 \be\label{sec2:bc}
 c={3\ell\over 2G_3}~.
 \ee
It is surprising that the addition of higher spin fields does not affect the central charge. The value in \eqref{sec2:bc} is the same as computed by Brown-Henneaux \cite{Brown:1986nw}  for Einstein gravity with a negative cosmological constant. This results also holds in the infinite $N$ limit, where the algebra is $\CW_\infty$ and the central charge is still \eqref{sec2:bc}  \cite{Henneaux:2010xg}.

The appearance of the centrally extended algebra as studied in \cite{Henneaux:2010xg,Campoleoni:2010zq}  is purely classical. The analysis presented in \cite{Gaberdiel:2010ar} goes one step further and tests whether the $\CW_N$ persists at the quantum level. These authors computed the 1-loop determinant associated to the linearized fluctuations \eqref{sec2:bd}.  They found that the full 1-loop contribution of a single spin $s$ field is simply
\be
Z^{(s)}=\prod_{n=s}^\infty|1-q^n|^{-2}~,
\ee
 where $q=\exp(2\pi i\tau)$ and $\tau$ is the complex structure of the torus at the boundary of thermal AdS$_3$. Therefore for a $SL(N)\times SL(N)$ Chern-Simons theory, which contains a family of spin fields from $s=2$ up to $s=N$, the 1-loop determinant is given by
\be\label{sec2:ca}
Z^{1-{\rm loop}}_{N}=\prod_{s=2}^{N} \prod_{n=s}^\infty|1-q^n|^{-2}=\chi_N\times\bar\chi_N~,
\ee
with
\be\label{sec2:aa}
\chi_N=\prod_{s=2}^N\prod_{n=s}^{\infty}(1-q^n)^{-1}~.
\ee
$\chi_N $ is precisely the vacuum character of the $\CW_N$ algebra. For infinite $N$ the resulting 1-loop determinant is
\be\label{sec2:cb}
Z^{1-{\rm loop}}_{\infty}=\prod_{s=2}^{\infty} \prod_{n=s}^\infty|1-q^n|^{-2}=\chi_\infty\times\bar\chi_\infty~,
\ee
where
\be\label{sec2:ab}
\chi_\infty=M(q)\prod_{n=1}^{\infty}(1-q^n)~,
\ee
and the MacMahon function is defined as
\be\label{sec2:ac}
M(q)=\prod_{n=1}^{\infty}(1-q^n)^{-n}~.
\ee
The function $\chi_\infty$ is the character of the $\CW_\infty$ algebra. One nice and unexpected feature is that equations \eqref{sec2:ca} and \eqref{sec2:cb} can be written as the square of a holomorphic function of $q$.  

Although these one loop determinants were computed directly in the bulk using heat kernel methods, in fact they have a simple physical interpretation.  They can be derived using strictly algebraic methods, as traces over the vacuum representations of $\CW_N$ and $\CW_\infty$.  This is the representation where all of the $\CW_N$ descendants are linearly independent and have positive norm; i.e. the representation without null vectors.   
Using this fact, it was further argued in \cite{Gaberdiel:2010ar} that the partition functions \eqref{sec2:ca} and \eqref{sec2:cb} are one-loop exact, following  \cite{Maloney:2007ud}.


\section{Partition function and growth of states}\label{sec:partition}

We would now like to study the general properties of the partition function of an asymptotically AdS theory of gravity with $\CW_N$ symmetry.  Our basic observation is that there is a tension between the two essential features described above -- the existence of asymptotic conformal symmetry with a finite central charge, and the appearance of the infinite tower of linearly independent, finite norm $\CW_N$ descendants.  In some cases these features are mutually incompatible.

We start by considering the partition function
\beq\label{sec3:aa}
Z(\tau,\bar\tau)=\sum_{\Delta,\bar\Delta}d(\Delta,{\bar\Delta})q^{\Delta}\bar q^{\bar\Delta}~,
\eeq
where $d(\Delta, {\bar \Delta})$ is the number of states with weight $(\Delta, {\bar \Delta})$.  We will use the conventional ``CFT normalization" for the weights so that the ground state (i.e. empty Anti-de Sitter space) has $\Delta={\bar \Delta}=-c/24$.  This partition function can be regarded as a Euclidean path integral in three dimensions, where we sum over all field configurations such that the metric approaches a torus at asymptotic infinity.\footnote{ The literature on the partition function of AdS$_3$ gravity is extensive, see e.g. \cite{Maldacena:1998bw,Witten:2007kt,Dijkgraaf:2000fq,Kraus:2006nb,Maloney:2007ud} and references therein.}  With standard Brown-Henneaux boundary conditions this partition function will be a function only of the conformal structure $\tau$ of the torus at infinity, and will hence be invariant under the modular transformation $\tau\to{-1/\tau}$.  In the gravitational language, this modular transformation is a large diffeomorphism of the bulk which induces a large conformal transformation of the boundary torus.  

Modular invariance leads to Cardy's formula \cite{Cardy:1986ie}
\be\label{sec3:ab}
\log( d(\Delta,\bar\Delta))\sim 2\pi\sqrt{c\Delta\over6} + 2\pi\sqrt{c\bar \Delta\over6}~,
\ee
for the number of states at large $\Delta, {\bar \Delta}$.  
The first assumption involved in the derivation of this formula is that the bulk theory is diffeomorphism invariant in Euclidean signature.  The second is that the ground state has finite norm, so that the first excited state has $\Delta, {\bar \Delta}>-c/24$ and is separated by a gap from the ground state.  Provided these assumptions are satisfied, equation  \eqref{sec3:ab} is universal.  The details of the bulk theory, such as the specific matter content, will only enter into the subleading corrections to this formula.

This universal behaviour is a consequence of the physics of AdS${}_3$ black holes.  Every classical theory of AdS${}_3$ gravity contains black holes \cite{Banados:1992wn,Banados:1992gq}.  These black holes are quotients of AdS${}_3$, so will necessarily exist as solutions to the equations of motion if AdS${}_3$ itself is a solution to the equations of motion.
Their Bekenstein-Hawking entropy is precisely given by equation \eqref{sec3:ab} \cite{Strominger:1997eq}.  Thus we expect that in a quantum theory of AdS${}_3$ gravity, there should be states with arbitrarily large weights which describe the BTZ black hole.

Let us now reconsider the higher spin theories in this light.  Although we will not be able to compute the partition function exactly, we can compute the tree and one-loop contributions.  The vacuum state will just be empty AdS, which contributes to the tree level partition function
\be
Z^{(0)} = q^{-c/24}\bar q^{-c/24}~.
\ee 
The one loop piece is also easy to compute.  It is given by the trace  
\be
Z^{(1)}={\rm Tr}_{\cal H} \left(q^{L_0} {\bar q}^{{\bar L}_0}\right)~,
\ee
over the Hilbert space ${\cal H}$ of linearized excitations of the theory.  This is the space of solutions to the linearized equations of motion \eqref{sec2:bb} modulo gauge transformations.  Since all local excitations are pure gauge, one might guess that there are no such contributions.  However, this is not quite the case as the set of allowed gauge transformations includes only those which vanish sufficiently quickly at infinity.  Thus the spectrum includes states obtained by acting on the vacuum state by a linearized gauge transformation at the boundary.  Indeed, it was argued that these gauge transformations generate the algebra $\CW_N$, which is an extension of the usual Virasoro algebra $\CW_2$.  
Thus the linearized fluctuations of the spin fields are organized into a $\CW_N$ character \cite{Gaberdiel:2010ar},  
\be\label{sec3:ac}
Z^{(1)}(q)=q^{-c/24}\bar q^{-c/24}|\chi_N(q)|^2~.
\ee
Here $\chi(q)$ is given by the vacuum character \eqref{sec2:aa} or \eqref{sec2:ab} depending on whether $N$ is finite or infinite.  

It is important to emphasize that there is nothing mysterious about the states which contribute to the partition function \eqref{sec3:ac}.  They describe solutions to the equations of motion and can be written out explicitly in the Chern-Simons language.  For $N=2$, of course, they have a simple interpretation; they are the usual boundary gravitons.  At the linearized level, these states have finite norm with respect to  Klein-Gordon inner product, so appear to represent legitimate states of the free higher spin field theory.  The question is whether these states will survive at the non-linear level, and if they do what the implications are for the quantum theory.

The most immediate effect of the higher spin fields is to increase the number of states at each level. In particular, the number of $\CW_N$ descendants of a given dimension is larger than the number of Virasoro descendants. But the total number of states is constrained by Cardy's formula \eqref{sec3:ab}.  If the linearized states appearing in \eqref{sec3:ac} are to appear as states in the full theory, this a significant constraint.

To see this let us first consider the case where $N$ is finite. The coefficients $p^N_\Delta$ of the $\CW_N$ vacuum character
\be\label{sec3:char}
\chi_N=\prod_{s=2}^N\prod_{n=s}^{\infty}(1-q^n)^{-1}=\sum_{\Delta} p^N_\Delta q^\Delta~,
\ee
can be estimated at large $\Delta$.  They grow like
\be\label{sec3:cb}
\log{(p^N_\Delta)}\sim 2\pi\sqrt{{(N-1)\over 6}\Delta}~,
\ee
when $\Delta$ is large (and in particular if $\Delta\gg N^3$).  A derivation of this is given in the appendix, but the origin of this growth can be understood intuitively.  When $N=2$, $\chi_2$ is the vacuum character of the Virasoro algebra.  In the absence of null vectors, the number of Virasoro descendants of a given primary state increases like the number of states in a CFT of central charge $c=1$.  That is why the construction CFTs with $c<1$ (the minimal models) is a highly constrained algebraic problem which requires the existence of null vectors.  For $N>2$, we observe that the character \eqref{sec3:char} is equal to the $(N-1)^{th}$ power of the Virasoro vacuum character times a finite polynomial in $q$.  Thus it is natural to guess that the number of descendants grows like the number of states of a CFT with central charge $(N-1)$.  From equation \eqref{sec3:cb}  we see that this is indeed the case.  One just has to verify that this finite polynomial does not lead to cancellations which will spoil this heuristic argument; this computation is described in  appendix \ref{AA:XN}.
 
Comparing equations \eqref{sec3:cb} and \eqref{sec3:ab} it is clear that if $N-1>c$ then there will be a value of $\Delta$ for which $p^N_\Delta$ will exceed the allowed density of states $d(\Delta,\bar \Delta)$.  Thus some of the linearized states must be removed from the spectrum.  Indeed, we will see explicitly that this can happen in certain cases in the next section for the bulk theories dual to the $\CW_N$ minimal models.  

We note that the situation is even more drastic if $N$ is infinite.  The descendants are counted by the $\CW_\infty$ character
\be
\chi_\infty=M(q)\prod_{n=1}^{\infty}(1-q^n)=\sum_{\Delta=1}^\infty p^\infty_\Delta q^\Delta~,
\ee
whose coefficients grow like
\be\label{sec3:ca}
\log{(p^\infty_\Delta)}\sim 3\left({\zeta(3)\Delta^2\over 4}\right)^{1/3}~,
\ee
as we show in appendix \ref{AA:MM}.  The growth of states in \eqref{sec3:ca} will always exceed the Cardy growth \eqref{sec3:ab} for any finite value of the central charge.  Thus in the absence of a linearization instability, the number of perturbative states vastly exceeds the number of black holes states.   

Finally, we  note that the convergence towards the asymptotic values \eqref{sec3:cb} and \eqref{sec3:ca} is rather slow.   In some cases, this might mean that in order to see that the number of $\CW_N$ descendants exceeds the Cardy bound we have to look at states of very high dimension.  

As an illustration of this phenomenon, we will consider the following simple example.  Let us ask if it is possible to construct a ``pure'' theory of gravity with $\CW_N$ symmetry, in the sense that the only perturbative states are the $\CW_N$ descendants described above.  Following \cite{Witten:2007kt}, it is natural to conjecture that this theory is holomorphically factorized.  In this case the partition function will be the square of an analytic function $Z(\tau)$ which diverges like $q^{-c/24}$ as $q\to 0$.  $Z(\tau)$ will be a holomorphic, modular invariant function on the upper half $\tau$ plane.  Using general properties of modular functions (see e.g. \cite{apostol}) it follows that $Z(\tau)$ is determined uniquely provided we specify the $c/24$ polar terms in the expansion of $Z(\tau)$ around $q=0$.  If the theory is ``pure'' in the sense defined above, then these polar terms are found by demanding that they match the polar terms in the one loop partition function \eqref{sec3:ac}.  It is then straightforward to compute $Z(\tau)$ for any desired values of $N$ and $c$ and hence determine the number of states of any dimension $\Delta$, using an algorithm similar to that presented in \cite{Witten:2007kt}.\footnote{
One could also compute $Z(\tau)$ by performing a sum over geometries, following \cite{Dijkgraaf:2000fq}.  If we simply sum the holomorphic part of the one-loop determinant over the coset $SL(2,\Z)/\Z$, then the resulting $Z(\tau)$ will be the same as that described above.  However, if one does not assume holomorphic factorization and instead sums the full one loop determinant \eqref{sec3:ac} over $SL(2,\Z)/\Z $ one finds results which are not consistent with a quantum mechanical interpretation, as in \cite{Maloney:2007ud}.}  
It is then possible to
check explicitly that for any $N-1>c$ there is some value of $\Delta$ for which the number of $\CW_N$ descendants exceeds to total number of states counted by the partition function $Z(\tau)$.

It is amusing to work this out explicitly for the case $c=24$ where the holomorphic part of the partition function is, up to an additive constant, equal to the Klein's $J$-invariant $J(\tau)$.  The $q$ expansion is
\be\label{qq}
Z(\tau)= q^{-1} + {(\rm const)} + 196884q + 21493760q^2+864299970 q^3+\dots
\ee
One can compare this to the asymptotic growth of the vacuum character 
\be\label{qqq}
Z^{(1)}(\tau) = q^{-1} \chi_N = q^{-1} + q + 2 q^2 + 3 q^3 \dots ~~.
\ee
It is a surprising (but true) fact that when $N>25$ the coefficients of \eqref{qqq} become larger than those of \eqref{qq} for some value of $\Delta$.  For $N$ very large this occurs when $\Delta\approx 10^5$ and the coefficients are of order $10^{1000}$.\footnote{In fact, we can improve this argument a bit by noticing that the full partition function must be a $\CW_N$ character, so that every time a primary state appears in the theory this leads to additional $\CW_N$ descendants at higher order.  When $N=\infty$ and $c=24$, for example, this leads to a negative number of $\CW_N$ primaries at $\Delta\approx 60000$ if there are no null vectors.} 
The explanation of this curiously large value of $\Delta$ is the following.  The $J$-function happens to be well approximated by Cardy's formula for small values of $\Delta$, whereas the corresponding asymptotic formula for $\chi_N$ is only a good approximation for relatively large (of order $10^4$) values of $\Delta$.  The lesson is that while the first few coefficients in expressions like \eqref{qqq} may appear small, this does not tell the full story!

\section{Minimal Models and Black Holes}

In this section we comment on the $\CW_N$ minimal models, which provide specific and calculable examples of $\CW_N$ symmetric CFTs with central charges $c<N-1$.  Thus they lie on the other side of the bound \eqref{eqq}. This bound was motivated in part by bulk gravity considerations, so one might expect that the bulk duals to these minimal models have several rather unusual properties.  Indeed we will see that they possess a linearization instability and that for finite $k$ and $N$ the spectrum of black hole states differs qualitatively from the semiclassical expectation.

The $\CW_N$ minimal model at level $k$ can be described in terms of the coset WZW model
\be
{{su}(N)_k\oplus {su}(N)_1\over {su}(N)_{k+1}}~,
\ee
where the subscripts give the level of the algebra.
The central charge  is   
\be
c = (N-1) \left(1- {N(N+1) \over (N+k)(N+k+1)}\right)~,
\ee
and is strictly less than $N-1$ for finite values of $N$ and $k$.
When $N=2$ these coincide with the usual (Virasoro) minimal models, and it can be proven that there are no other unitary CFTs with $c<1$. 
We do not know of a similar proof for higher values of $N$.

We note that, from a quantum gravity perspective, these higher $N$ minimal models are much more interesting than their Virasoro ($c<1$) cousins.  That is because $c$ can be taken to be large provided that $N$ is also large, so that the theories are dual to macroscopic theories of three dimensional gravity with AdS radius large in Planck units.  Thus one would expect all of the familiar features of classical three dimensional gravity -- in particular the BTZ black holes -- to arise in this limit.  

Unfortunately, the bulk duals of these theories are not known explicitly.  However, when $N$ and $k$ are taken to infinity with the ratio $k/N$ fixed, the bulk dual was conjectured to be an infinite tower of higher spin fields along with a pair of complex scalar fields  \cite{Gaberdiel:2010pz}.  In this limit the central charge goes to infinity, meaning that the AdS radius is infinite in Planck units.  For finite values of $N$ this bulk theory should presumably be augmented by terms involving the curvature of AdS space.  These modifications are not known, but we can still describe some basic features of the bulk dual of the $\CW_N$ minimal models for finite $N$ and $k$.  

As emphasized in the previous section, the theory must have null vectors, meaning that certain higher spin versions of the boundary gravitons are  removed from the spectrum.   Indeed, one can check explicitly that the $\CW_N$ minimal models have null vectors.  For the $\CW_N$ descendants of the vacuum, the first null state appears at dimension $\Delta =k+1-{c\over 24}$.  Indeed, the vast majority of the $\CW_N$ descendants will be projected out of the spectrum at high order.

In fact, for large values of $\Delta$ the spectrum of the $\CW_N$ minimal model consists $entirely$ of descendant states, rather than primary states.  In particular, these theories have only a finite number of primaries, hence they have a state with largest dimension.  The dimension of this highest dimension state can be estimated, and is of order\footnote{We are grateful to M. Gaberdiel for discussions related to this point.}
\be\label{w}
\Delta_{\rm max} \sim k^2 N~,
\ee
when $k$ and $N$ are large.

It is worth commenting on the bulk interpretation of these descendant states.  The only states with arbitrarily high dimension (above $\Delta_{\rm max}$) are the descendants of lower dimension primary states.
Thus in the bulk the high energy spectrum consists entirely of lower energy states which are dressed by a large number of boundary excitations.  The exponentially large degeneracy of states at high energy comes from the large number of such descendant states -- i.e. from the large number higher spin boundary excitations.  

We note that this is in drastic contrast to our semiclassical expectations.  BTZ black holes exist as classical solutions of the equations of motion for any value of the mass and angular momentum such that $M\ell \ge  J$.  In particular the theory contains black holes whose horizon size ($r_+$) is arbitrarily large compared to both the Planck length $G_3$ and the AdS radius $\ell$.  In the dual CFT language, these correspond to states with dimensions $\Delta$ such that 
\be\label{sec4:ba}
c\Delta\sim \left({r_+^2 \over G^2_3 }\right)\gg 1~,
\ee
and 
\be\label{sec4:bb}
\left({\Delta\over c}\right) \sim \left({r_+^2 \over \ell^2 }\right)\gg 1~.
\ee
For states where the second inequality is valid, Cardy's formula can be used to compute the entropy.  Although we might expect that the allowed values $r_+$ (and hence $\Delta$) will be quantized in the full quantum theory, we still expect that there should still be a tower of black hole states with arbitrarily large dimension.

However, this is not what is indicated by equation \eqref{w} for finite values of $k$ and $N$.  The states with $\Delta\gg\Delta_{\rm max}$ are  lower energy states dressed by a large number of perturbative excitations.  Thus, if they are to be interpreted as black holes, they should be regarded as a black hole of small area (a primary with $\Delta \le \Delta_{\rm max}$) along with  some number of boundary excitations.  When $\Delta \gg \Delta_{\rm max}$ most of the energy (and  entropy) of a state in the high energy spectrum comes just from these boundary excitations.  In this sense the bulk dual of a minimal model does not appear to possess black holes with arbitrarily large mass and angular momentum.

We emphasize that this picture may be altered in the large $N$ or $k$ limit.  Depending on how this limit is approached the minimal ${\CW}_N$ models may contain large black holes in the sense of \eqref{sec4:ba} and \eqref{sec4:bb}.   For example, in the 't Hooft limit 
(as defined in  \cite{Gaberdiel:2010pz}) $N$ and $k$ are taken to infinity with the ratio $k/N$ fixed.  This is a classical limit in the bulk where the central charge becomes infinite.   For the highest dimension state in the theory, the ratios \eqref{sec4:ba} and \eqref{sec4:bb} become large.  A different, and somewhat simpler,  case to consider is $k$ large with $N$ fixed.  In this limit $c\to (N-1)$ and the linearization instability (i.e. the null vectors) disappears.  Again the ratios  \eqref{sec4:ba} and \eqref{sec4:bb} become large. So in both cases there are primary states of large dimension which might be interpreted as BTZ microstates.   It would be interesting to compute the degeneracies of these states and see if they can indeed be interpreted as black holes (see \cite{Kiritsis:2010xc} for related progress in this direction).

\section*{Acknowledgements}

We are grateful to R. Gopakumar, E. Perlmutter and especially M. Gaberdiel for interesting discussions and useful explanations.  This work was supported by the National Science and Engineering Research Council of Canada and FQRNT (Fonds qu\'eb\'ecois de la recherche sur la nature et les technologies).

\newpage

\appendix

\section{Asymptotic Behaviour of $\chi_N$ and $\chi_\infty$}
\label{sec:AA}

Here we collect several asymptotic formulas for $\CW_N$ characters used in section  \ref{sec:partition} and sketch the derivations of these formulas.

\subsection{Asymptotics of the Partition Function $F(q)$}\label{sec:A1}  

As a warmup we first estimate the growth of the coefficients of partition function $F(q)$ defined as
\be\label{AA:a1}
F(q)=\sum_{n=0}^{\infty}p(n)q^{n}=\prod_{n=1}^\infty(1-q^n)^{-1}~.
\ee
where $p(n)$ is the number of partitions of the integer $n$.
Our goal is to approximate $p(n)$ for large values of $n$.  This is a classic computation which we review here for the sake of completeness.

We start with the inverse Laplace transform
\be\label{AA:a}
p(n)={1\over 2\pi i}\int_{\cal C} {F(q)\over q^{n+1}}dq~,
\ee
where $\cal C$ is a simple contour that encloses the origin.  Since $F(q)$ has poles for $|q|=1$ we must keep the contour $\cal C$ inside the unit circle in the complex $q$ plane. Our strategy is to choose a contour $\cal C$ which approaches the unit circle $|q|=1$ where we can approximate $F(q)$ by elementary functions.

Our next step is to write $F(q)$ as  an elliptic modular function
\be 
F(e^{2\pi i\tau})=e^{i\pi\tau/ 12}\eta(\tau)^{-1}~,
\ee
with $\eta(\tau)$ the Dedekind eta function and $q=e^{2\pi i\tau}$.  The eta function transforms simply under modular transformations.  In particular, 
\be
\eta\left(-{1/\tau}\right) = (-i \tau)^{1/2} \eta(\tau)~, 
\ee
%
%
%
so that
\be\label{AA:aa}
F(e^{2\pi i\tau})=\exp\left({i\pi\over 12}(\tau+\tau^{-1})\right)(-i\tau)^{1/2}F(e^{-2\pi i/\tau})~.
\ee
When the imaginary part of $\tau$ is very small, so that we are close to $|q|=1$, $F(e^{-2\pi i/\tau})$ approaches one and equation \eqref{AA:aa} becomes
\be\label{AA:b}
F(e^{2\pi i\tau})\sim \exp\left(i\pi(\tau+\tau^{-1})/12\right)(-i\tau)^{1/2}~,
\ee
so that
\be
p(n)\sim \int_{i\epsilon}^{i\epsilon+1} \exp\left(-2\pi i\tau(n-{1\over 24})+{\pi i\over 12\tau}\right)(-i\tau)^{1/2} d\tau~.
\ee
Using the saddle point approximation we obtain
\be
p(n)\sim ({\rm const}) {1\over n}\exp\left(\pi\sqrt{2n\over3}\right)~.
\ee
This estimate is valid only in the limit $n\to\infty$.  By refining the above argument we can estimate the size of the error terms in this approximation (see e.g. \cite{apostol}).

\subsection{Asymptotics of the $\CW_N$ character $\chi_N$}\label{AA:XN}

We now turn to the vacuum character for $\CW_N$, 
\be\label{AC:a}
\chi_N=\left(\prod_{n=1}^{N-1}(1-q^n)^{N-n} \right) F(q)^{N-1}=\sum_{n=0}^\infty p^N_n q^n ~. 
\ee
whose coefficients are again given by the contour integral
\be\label{AC:aa}
p^N_n=\int_{\cal C}{\chi_N\over q^{n+1}}dq~.
\ee
Again,  it is necessary to keep the contour within the unit circle $|q|=1$, where $\chi_N$ diverges.  Our goal is to obtain an approximate expression for $p_n^N$ by estimating $\chi_N$ when $|q| \to 1$. 

We start by noting that $\chi_N$ differs from $F(q)^{N-1}$ only by the prefactor in parenthesis in \eqref{AC:a}.
We then define the log of the polynomial prefactor in \eqref{AC:a}
\beq\label{AC:ab}
g(z)&=&\sum_{n=1}^{N-1}(N-n)\log(1-q^n)
=-\sum_{m=1}^\infty\sum_{n=1}^{N-1}(N-n){e^{-2\pi z nm}\over m}
\eeq
where $q=e^{-2\pi z}$. Our strategy will be to apply the Abel-Plana formula
\be\label{AB:a}
\sum_{n=0}^{\infty} f(n)=\int_0^{\infty}f(x)dx+{1\over2}f(0)+i\int^\infty_0{f(ix)-f(-ix)\over e^{2\pi x}-1}dx~,
\ee
which relates an infinite sum to the residues of a complex function. 
To use this formula we will first take a derivate of \eqref{AC:ab} and add the $m=0$ contribution
\beq
g'(z)=2\pi\sum_{m=0}^\infty\sum_{n=1}^{N-1}(N-n)n{e^{-2\pi z nm}}-{\pi\over 3}N(N^2-1)~.
\eeq
so that \eqref{AB:a} gives
\beq\label{AC:ba}
g'(z)&=&{1\over 2z}N(N-1)-{\pi\over 6}N(N^2-1)+{\cal O}(z)
\eeq
where we neglect terms which vanish in the $z\to0$ limit, where $|q|\to 1$.
%
%
%
%
Thus
\be\label{AC:bb}
g(z)={N(N-1)\over 2}\log(z)+ g_0-{\pi\over 6}N(N^2-1)z+{\cal O}(z^2)
\ee
where we have introduced a constant of integration $g_0$.

\begin{figure}[h!]
    \begin{center}
      \resizebox{90mm}{!}{\includegraphics{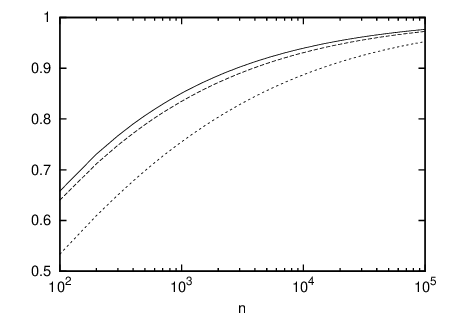}}
      \end{center}
      \caption{For $N=2$ (straight line), $N=3$ (long dash) and $N=6$ (short dash), we plot the ratio of numerical value of $\log(p_n^N)$ over the approximated value given by \eqref{AC:da}. As $N$ increases, we require larger values of $n$ to reach the Cardy regime. }
      \label{Chi3Figure}
  \end{figure}

Following the arguments of section \ref{sec:A1}, we approximate the contour integral by the value of $\chi_N$ close to $|q|=1$ where
\be\label{AC:ca}
\chi_N\sim z^{{1 \over 2}(N^2-1)}\exp\left(-{\pi\over 6}N(N^2-1)z-{\pi\over12}(N-1)(z-z^{-1})\right)~.
\ee
We have neglected an overall constant prefactor.
Thus %
\be
p^N_n\sim i\int_{\cal C}  z^{{1 \over 2}(N^2-1)}\exp\left(-{\pi\over 6}N(N^2-1)z-{\pi\over12}(N-1)(z-z^{-1})+2\pi nz\right)dz~.
\ee
and the saddle point approximation 
\be\label{AC:da}
p^N_n\sim ({\rm const})\, n^{-{1\over 4}(N^2+2)}
\exp\left(\pi\sqrt{2(N-1)n\over 3}\right) ~.
\ee
gives an estimate for $p^N_n$ which is valid in the large $n$ limit.
We note that the constant multiplying \eqref{AC:da} depends on $N$. In the figure\eqref{Chi3Figure} we compare the asymptotic formula \eqref{AC:da} with the actual values \eqref{AC:a}.  We note that the $p_n^N$ approach their asymptotic values more slowly as $N$ increases.  Indeed one can check that the error terms in this approximation are negligible only when $n\gg N^3$. 
      
\subsection{Asymptotics of the $\CW_\infty$ character $\chi_\infty$ and the MacMahon Function}\label{AA:MM}

For the character of $\CW_\infty$ we need to be a little more careful; we refer the reader to \cite{Wright} for a more detailed analysis. 
 The character is given by
\be\label{AB:d}
\chi_\infty=M(q)F(q)^{-1}~,
\ee
with $F(q)$ given by \eqref{AA:a1} and $M(q)$ the MacMahon function
  \be
  M(q) = \prod_{n=1}^\infty (1-q^n)^{-n}
  \ee
As before we compute the coefficients of $\chi_\infty$ using a contour integral
\be
p^\infty_n={1\over 2\pi i}\int_{\cal C} {\chi_\infty(q)\over q^{n+1}}dq~.
\ee
where $\cal C$ encloses the origin and is contained in the unit circle. 

We start by approximating the MacMahon function $M(q)$.  Defining the logarithm 
\be\label{AB:b}
g(z)\equiv\log M(q)=-\sum_{n=1}^{\infty} n\log(1-e^{-2 \pi n z})~.
\ee
with $q=e^{-2\pi z}$ 
%
and applying \eqref{AB:a} we find
\beq\label{AB:ca}
g(z)&=&-\int_0^\infty x\log(1-e^{-2\pi zx})dx+2\int_0^\infty{x\over e^{2\pi x}-1}\log(2\sin(\pi z x))dx\cr
&=&{\zeta(3)\over 4\pi^2 z^2}+{1\over 12}\log z + {1\over 12}(1-\gamma+6\zeta'(2))+{\cal O}(z^2)
\eeq
where we have used the Taylor expansion of $\log(2 sin(\pi z x))$ at $z\to 0$ and computed the integrals explicitly.
%
%
%
%
%
From this we can read off the behaviour of $M(q)$ at small $z$
\be
M(e^{-2\pi z})\sim z^{1/12}\exp\left({\zeta(3)\over 4\pi^2 z^2}\right)~.
\ee
where we have neglected an overall constant prefactor.
 
\begin{figure}[h]
    \begin{center}
      \resizebox{80mm}{!}{\includegraphics{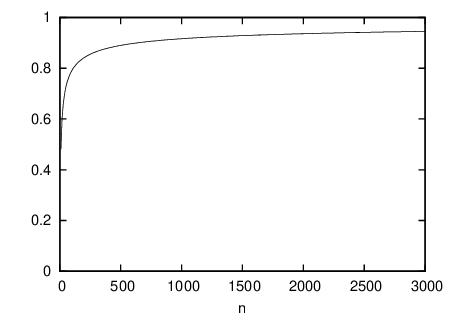}}
       \end{center}
      \caption{Ratio of $\log p^\infty_n$ over  the saddle point approximation  \eqref{AB:z}.}      
      \label{ChiMFigure}
  \end{figure}

This leads to an approximate expression for $\chi_\infty$
\be
\chi_\infty (e^{-2\pi z})\sim z^{-5/12}\exp\left({\zeta(3)\over 4\pi^2z^2}+{\pi\over 12}(z-z^{-1})\right)~.
\ee
so that
\be
p_n^{\infty}=i\int_{\cal C}  z^{-5/12}\exp\left({\zeta(3)\over 4\pi^2z^2}+{\pi\over 12}(z-z^{-1})+2\pi nz\right) dz~,
\ee
and the saddle point approximation gives
\be\label{AB:z}
p_n^\infty\sim({\rm const})~ n^{-19/36}\exp\left(3\left(\frac{\zeta(3)}{4}n^{2}\right)^{1/3}\left[1-\frac{\pi^{2}}{18}\left(\frac{2}{\zeta(3)^{2}n}\right)^{1/3}\right]\right)
~.
\ee
for large $n$.  We note that this grows like $e^{n^{2/3}}$, which is faster than the ${e^{ n^{1/2}}}$ behaviour obtained for finite $N$.
In figure \eqref{ChiMFigure} we compare the asymptotic growth \eqref{AB:z} with the actual coefficients of \eqref{AB:d}.  
 
 \newpage

\bibliographystyle{utphys}
\bibliography{all}

\end{document}